%% file: robert.tex
\documentclass[11pt]{article}
\usepackage[T1]{fontenc}
\usepackage{amsfonts}
\usepackage{amsmath}
\usepackage{amssymb}
\usepackage{geometry}
\usepackage{graphics}
\usepackage{color}

\newtheorem{lemma}{Lemma}
\newtheorem{theorem}{Theorem}
\newtheorem{corollary}{Corollary}

\newtheorem{definition}{Definition}
\newtheorem{remark}{Remark}

\def\cqfd{\hfill$\square$}
\def\B{\{0,1\}}

\def\1n{1,\dots,n}
\def\cst{\mathrm{cst}}

\def\s{\sigma}
\def\case#1{\left\{\begin{array}{#1}}
\def\esac{\end{array}\right.}
\def\x{\tilde x}

\def\F{\tilde F}
\def\f{\tilde f}

\title{On the length of attractors in boolean networks\\ with an interaction graph by layers.}

\author{
Adrien Richard\\[3mm] 
{\small{Laboratoire I3S}}\\ 
{\small{CNRS \& Université de Nice-Sophia Antipolis}}\\[3mm]
{\small{\texttt{richard@unice.fr}}}
}
\date{}

\begin{document}

\maketitle

\begin{abstract}
We consider a boolean network whose interaction graph has no circuit
of length $\geq 2$. Under this hypothesis, we establish an upper bound on
the length of the attractors of the network which only depends
on its interaction graph.
\end{abstract}

\section{Introduction}

We consider a boolean network $F:\{0,1\}^n\to\{0,1\}^n$ and its
interaction graph $G(F)$. The vertices correspond to the components of
the network, and there is a positive (resp. negative) edge from $j$ to $i$
if the component $j$ has a positive (resp. negative) effect on the
component~$i$. Then, under the assumption that $G(F)$ has no circuit of length
$>1$ (directed graphs without circuit of length $\geq 2$ are called graph
by layers in {\cite{GS07}}), we establish an
upper bound on the length of the attractor of the network which only
depends on $G(F)$. This result is related to a recent work of Goles and
Salinas {\cite{GS07}}.

\section{Definitions}

Let $n$ be a positive integer, and let $F$ be a map from $\B^n$ to
itself:
\[
x=(x_1,\dots,x_n)\in\B^n~\mapsto~F(x)=(f_1(x),\dots,f_n(x))\in\B^n.
\]
As usual, we see $F$ has as a synchronous boolean network with $n$
components: when the network is in state $x$ at time $t$, it
is in state $F(x)$ at time $t+1$.\\

A {\emph{path}} of $F$ of length $r\geq 1$, is a sequence
$(x^0,x^1,\dots,x^r)$ of points of $\B^n$ such that $F(x^k)=x^{k+1}$
for all $0\leq k<r$. A {\emph{cycle}} of $F$ of length $r\geq 1$ is a
path $(x^0,x^1,\dots,x^r)$ such that $x^0=x^r$ and such that the
points $x^0,\dots,x^{k-1}$ are pairwise distinct. The cycles of $F$
correspond to the attractors of the network.\\

We set $\bar 0=1$ and $\bar 1=0$. Then, for all $x\in\B$, we denote by
$\bar x^i$ the points $y$ of $\B^n$ defined by $y_i=\bar x_i$ and
$y_j=x_j$ for all $j\neq i$. For all $x\in \B^n$, we set:
\[
f_{ij}(x)=\frac{f_i(\bar x^j)-f_i(x)}{\bar x_j-x_j}\qquad (i,j=\1n).
\]
$f_{ij}$ may be see has the partial derivative of $f_i$ with
respect to the variable $x_j$.\\

We are now in position to define the interaction graph of the network:
the {\emph{interaction graph of $F$}}, denoted $G(F)$, is the graph
whose set of vertices is $\{\1n\}$ and which contains an edge from $j$
to $i$ of sign $s\in\{-1,1\}$ if there exists $x\in \B^n$ such that
$s=f_{ij}(x)$. So each edge of $G(F)$ is directed and labelled with a
sign, and $G(F)$ can contains both a positive and a negative edge from
one vertex to another. Note that there exists an edge from $j$ to $i$
in $G(F)$ if and only if $f_i$ depends on $x_j$.\\

Let $i,j$ be two vertices of $G(F)$. We say that $i$ is a
{\emph{successor}} (resp. {\emph{predecessor}}) of $j$ if $G(F)$ has
an edge from $j$ to $i$ (resp. from $i$ to $j$). We say that $i$ is a
{\emph{strict successor}} (resp. {\emph{strict predecessor}}) of $j$
if $i$ is a successor (resp. predecessor) of $j$ and $i\neq j$. A path
of $G(F)$ of length $r\geq 0$ is a sequence $P=(i_0,\dots,i_r)$ of
vertices of $G(F)$ such that $i_{k+1}$ is a successor of $i_{k}$ for
all $0\leq k<r$. We say that $P$ is a path from $i_0$ to $i_r$, and
that $P$ is {\emph{elementary}} if the vertices $i_0,\dots,i_{r}$
are pairwise distinct. A {\emph{circuit}} of $G(F)$ of length $r\geq
1$ is a path $(i_0,\dots,i_r)$ such that $i_0=i_r$ and such that the
vertices $i_0,\dots,i_{r-1}$ are pairwise distinct. A positive
(resp. negative) edge from a vertex $i$ to itself is called a
{\emph{positive (resp. negative) loop}} on $i$.

\begin{definition}
Let $P$ be an elementary path of $G(F)$. We denote by $\tau_{G(F)}(P)$
the number of vertices $i$ in $P$ satisfying at least one of the two
following properties: 
\begin{enumerate}
\item
$i$ is the first vertex of $P$ with a negative loop;
\item
 $i$ has both a positive and a negative
loop.
\end{enumerate}
We set 
$\tau(G(F))=\max\{\tau_{G(F)}(P),~\textrm{$P$ is an elementary path of $G(F)$}\}$.
\end{definition}

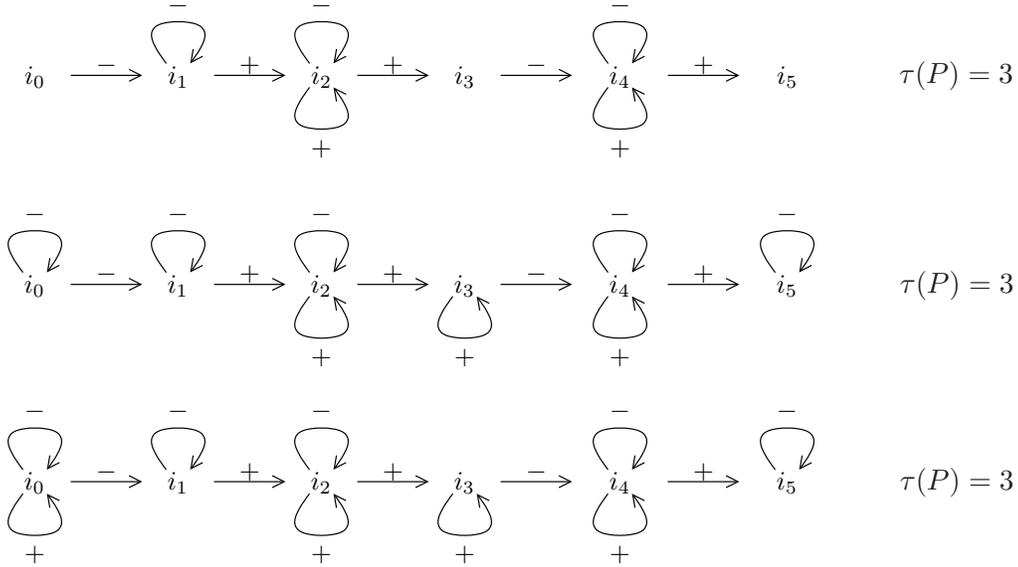
\begin{figure}
\[
\begin{array}{c}
\input{ex1.pstex_t}
\end{array}\qquad
\begin{array}{c}
\tau(P)=3\\[2mm]
\end{array}
\]
\[
\begin{array}{c}
\input{ex2.pstex_t}
\end{array}\qquad
\begin{array}{c}
\tau(P)=3\\[2mm]
\end{array}
\]
\[
\begin{array}{c}
\input{ex3.pstex_t}
\end{array}\qquad
\begin{array}{c}
\tau(P)=3\\[2mm]
\end{array}
\]
{\caption{\label{fig} Illustration of Definition 1.}}
\end{figure}

See Figure~{\ref{fig}} for an illustration of this definition. Note
that $\tau(G(F))\geq 1$ if and only if $G(F)$ has a negative loop, and
that $\tau(G(F))\leq 1$ if there is no vertex with both a positive and
a negative loop.

\section{Result}

Goles and Salinas {\cite{GS07}} proved the following theorem:

\begin{theorem}
Let $F:\B^n\to\B^n$ be such that $G(F)$ has no circuit
of length $\geq 2$. If $F$ has a cycle, then the length of this cycle if a
power of two, and it is $1$ if $G(F)$ has no negative loop.
\end{theorem}

The aim of this note is to prove the following extension:

\begin{theorem}
Let $F:\B^n\to\B^n$ be such that $G(F)$ has no circuit
of length $\geq 2$. If $F$ has a cycle, then the length of this cycle is a
power of two less than or equal to $2^{\tau(G(F))}$.
\end{theorem}

The proof needs few additional definitions. Let $F$ and $\F$ be two
maps from $\B^n$ to itself. We say that $G(\F)$ is a {\emph{subgraph}}
(resp. a {\emph{strict subgraph}}) of $G(F)$ if the set of edges of
$G(\F)$ is a subset (resp. a strict subset) of the set of edges of
$G(F)$. We say that $F$ is {\emph{$r$-minimal}} if $F$ has a cycle of
length $r$ and if there is no map $\F$ with a cycle of length $r$ such
that $G(\F)$ is a strict subgraph of $G(F)$. Note that if $F$ has a
cycle of length $r$, there always exists a $r$-minimal map $\F$ such
that $G(\F)$ is a subgraph of $G(F)$.

\begin{lemma}
Let $F:\B^n\to\B^n$ be $r$-minimal with $r\geq 2$, and assume that $G(F)$ has no circuit of length $\geq 2$. There exists a map $\F:\B^n\to\B^n$ with
a cycle of length $r/2$ such that $G(\F)$ is a subgraph of $G(F)$ and
such that $\tau(G(\F))<\tau(G(F))$.
\end{lemma}

\noindent 
{\bf{Proof $-$}} Let $\s=(x^0,\dots,x^r)$ be a cycle of $F$ of
length $r$. To simplify notations, we set $x^{k+r}=x^k$ for all
positive integer $k$. Since $\s$ is of length $\geq 2$, $F$ is not
constant. Thus, there exists a vertex $j$ in $G(F)$ with a predecessor. Let $P$
be an elementary path of $G(F)$ of maximal length starting from $j$,
and let $i$ be the last vertex of this path. Then:
\[
{\textrm{\emph{The vertex $i$ has a predecessor and no strict successor in $G(F)$}}}.
\]
The fact that $i$ has a predecessor is obvious if $i\neq j$, and true
by hypothesis if $i=j$; $i$ has no strict successor since if not, the
path $P$ being elementary and of maximal length, $G(F)$ would have a
circuit of length $\geq 2$.\\

Let $\F:\B^n\to\B^n$ be defined by:
\[
\f_i=\cst=0,\qquad \f_j=f_j\quad\textrm{for all}\quad j\neq i. 
\]
It is easy to see that $G(\F)$ is the subgraph of $G(F)$ that we
obtain by removing all the edges whose end vertex is $i$. Since $i$
has a predecessor in $G(F)$, we deduce that:
\[
\textrm{\emph{$G(\F)$ is a
{\emph{strict}} subgraph of $G(F)$}}.
\]
In the following, we prove that $\F$ has a cycle of length $r/2$ and
that $\tau(G(\F))<\tau(G(F))$.\\

For all integer $k$, let $\x^k$ be the point of $\B^n$ defined by:
\[
\x^k_i=0,\qquad \x^k_j=x^k_j\quad\textrm{for all}\quad j\neq i. 
\]
Since $\f_j=f_j$ does not depend on $x_i$ for all $j\neq i$ (vertex
$i$ has no strict successor in $G(F)$), and since $\f_i=\cst$ does not
depend on $x_i$, we have $\F(\x^k)=\F(x^k)$ and we deduce that:
\[
\F(\x^k)=\F(x^k)=(0,f_2(x^k),\dots,f_n(x^k))=(0,x^{k+1}_2,\dots,x^{k+1}_n)=\x^{k+1}.
\]
In other words, $(\x^0,\dots,\x^r)$ is a path of $\F$. Since
$\x^0=\x^r$, we deduce that $\F$ has a cycle $(\x^0,\dots,\x^p)$ of
length $p\leq r$. Then, for all integer $k$, we have:
\begin{equation}\label{eq:tilde}
\x^{k+p}=\x^k.
\end{equation}
Since $G(\F)$ is a strict subgraph of $G(F)$, and since $F$ is
$r$-minimal, we have $p<r$. Consequently, for all integer $k$:
\[
x^{k+p}\neq x^k.
\]
From this and (\ref{eq:tilde}), we deduce that, for all integer $k$:
\begin{equation}\label{eq:bar}
x^{k+p}=\overline{x^k}^i.
\end{equation}
Consequently, 
\[
x^{k+2p}=\overline{x^{k+p}}^i=\overline{\overline{x^k}^i}^i=x^k
\]
and we deduce that $2p=r$: $\F$ has indeed a cycle of length
$r/2$.\\

Let $j$ be any vertex of $G(F)$ with a predecessor and without strict
successor. With similar argument, we can show that $x^{k+r/2}=\overline{x^k}^j$. Then
$\overline{x^k}^j=\overline{x^k}^i$ so that $i=j$. Consequently:
\begin{equation}\label{eq:unique}
\begin{array}{c}
{\textrm{\emph{The vertex $i$ is the {\emph{unique}} vertex of
$G(F)$}}}\\{\textrm{\emph{ with a predecessor and without
strict successor.}}}
\end{array}
\end{equation}
We deduce that:
\begin{equation}\label{eq:path}
\begin{array}{c}
{\textrm{\emph{If a vertex $j$ has a predecessor in $G(F)$, then $G(F)$ has a path from $j$ to $i$.}}}
\end{array}
\end{equation}
Indeed, let $j$ be a vertex with a predecessor, and let $P$ an
elementary path of $G(F)$ of maximal length starting from $j$. As
argued above, the last vertex of $P$ has a predecessor and no strict
successor. We then deduce from (\ref{eq:unique}) that the last vertex of
$P$ is $i$.\\

Now, we prove that:
\begin{equation}\label{eq:loop1}
\begin{array}{c}
{\textrm{\emph{The vertex $i$ has a negative loop in
$G(F)$.}}}
\end{array}
\end{equation}
Since $x^p=\overline{x^0}^i$, we have $x^0_i\neq x^p_i$, and we deduce
that there exists $0\leq k<p$ such that:
\[
x^k_i\neq x^{k+1}_i.
\]
Then:
\[
f_i(x^k)=x^{k+1}_i=\overline{x^k_i}=x^{k+p}_i.
\]
Moreover, we have 
\[
x^{k+1+p}_i\neq x^{k+1}_i
\]
so
\[
f_i(x^{k+p})=x^{k+p+1}_i=\overline{x^{k+1}_i}=x^k_i
\]
and using (\ref{eq:bar}) we deduce that:
\[
f_{ii}(x^k)=
\frac{f_i(x^{k+p})-f_i(x^k)}{x^{k+p}_i-x^k_i}=
\frac{x^k_i-x^{k+p}_i}{x^{k+p}_i-x^k_i}=-1.
\]~

In addition:
\begin{equation}\label{eq:loop2}
\begin{array}{c}
{\textrm{\emph{If $i$ has a strict predecessor in $G(F)$, then $i$ has a
positive loop in $G(F)$.}}}
\end{array}
\end{equation}
Suppose that $i$ has a strict predecessor, and suppose that $x^k_i\neq
x^{k+1}_i$ for all $k$. Consider the map $\bar F:\B^n\to\B^n$ defined
by $\bar f_i(x)=\bar x_i$ and $\bar f_j=f_j$ for $j\neq i$. Clearly,
$\sigma$ is a cycle of $\bar F$, and $G(\bar F)$ is the subgraph of
$G(F)$ that we obtain by removing the edges whose end vertex is $i$,
expect the negative loop on $i$ (whose existence is proved). Since $i$
has a strict predecessor in $G(F)$, we deduce that $G(\bar F)$ is a
strict subgraph of $G(F)$, and this is not possible since $F$ is
$r$-minimal. Thus there exists $k$ such that
\[
x^k_i= x^{k+1}_i=f_i(x^k).
\]
Then
\[
x^{k+p}_i\neq x^k_i=x^{k+1}_i
\qquad\textrm{and}\qquad
x^{k+p+1}_i\neq x^{k+1}_i
\]
so
\[
x^{k+p}_i= x^{k+p+1}_i=f_i(x^{k+p})
\]
and using (\ref{eq:bar}) we deduce that:
\[
f_{ii}(x^k)=
\frac{f_i(x^{k+p})-f_i(x^k)}{x^{k+p}_i-x^k_i}=
\frac{x^{k+p}_i-x^k_i}{x^{k+p}_i-x^k_i}=1.
\]~

We are now in position to prove that $\tau(G(\F))<\tau(G(F))$. Since
$i$ has a negative loop in $G(F)$, we have $\tau(G(F))>0$. So suppose
that $\tau(G(\F))>0$, and let $P$ be an elementary path of $G(\F)$
such that
\[
\tau_{G(\F)}(P)=\tau(G(\F)).
\]
Since $G(\F)$ is a subgraph of $G(F)$, $P$ is an elementary path of
$G(F)$ and
\[
\tau_{G(\F)}(P)\leq \tau_{G(F)}(P).
\]
Let $j$ be the first vertex of $P$ with a negative loop in $G(\F)$
($j$ exists since $\tau(G(\F))>0$), and let $k$ be the last vertex of
$P$. Then $k$ has a predecessor in $G(\F)$ (this is obvious if $k\neq
j$ and also true if $k=j$ since $j$ has a negative loop) and thus
$k\neq i$ (since $i$ has no predecessor in $G(\F)$). So $k$ has a
predecessor in $G(F)$ and following (\ref{eq:path}), there exists an
elementary path $P'$ from $k$ to $i$ in $G(F)$. Since $G(F)$ has no
circuit of length $\geq 2$, the concatenation $Q$ of $P$ and $P'$ is
an elementary path of $G(F)$, and since $k\neq i$, $i$ has a strict
predecessor~in~$G(F)$. We then deduce from (\ref{eq:loop1}) and
(\ref{eq:loop2}) that $i$ has both a positive and a negative loop in
$G(F)$. It is then clear that
\[
\tau(G(\F))\leq \tau_{G(F)}(P)<\tau_{G(F)}(Q)\leq \tau(G(F))
\]
\cqfd\\

\noindent
{\bf{Proof of Theorem 2 $-$}} Let $F:\B^n\to\B^n$ be such that $G(F)$
has no circuit of length $\geq 2$ and suppose that $F$ has a cycle of
length $r$. We want to prove that $r$ is a power of two less than or
equal to $2^{\tau(G(F))}$. We proceed by induction on $r$. The base
case $r=1$ is obvious. So suppose that $r>1$. The induction hypothesis
is:
\begin{quote}
Let $\F:\B^n\to\B^n$ be such that $G(\F)$ has no circuit of length
$\geq 2$.\\ If $\F$ has a cycle of length $l<r$, then $l$ is a power of two
$\leq 2^{\tau(G(\F))}$.
\end{quote}
Consider a
$r$-minimal map $\bar F:\B^n\to\B^n$ such that $G(\bar F)$ is a
subgraph of $G(F)$. Then $G(\bar F)$ has no circuit of length $\geq 2$, and
following Lemma 1, there exists a map $\F$ with a cycle of length
$r/2$ such that $G(\F)$ is a subgraph of $G(F)$ and such that
$\tau(G(\F))<\tau(G(F))$. Since $G(\F)$ is a subgraph $G(\bar F)$,
$G(\F)$ has no circuit of length $\geq 2$. So, by induction hypothesis,
$r/2$ is a power of two $\leq 2^{\tau(G(\F))}$. So $r$ is a power of
two, and since $\tau(G(\F))<\tau(G(\bar F))$ we have $r\leq
2^{\tau(G(\bar F))}$. Since $G(\bar F)$ is a subgraph of $G(F)$, we
have $\tau(G(\bar F))\leq\tau(G(F))$ and we deduce that $r\leq
2^{\tau(G(F))}$.\cqfd\\

Let us say that $G(F)$ has an {\emph{ambiguous loop}}, if $G(F)$ has a
vertex with both a positive and a negative loop.

\begin{corollary}
Let $F:\B^n\to\B^n$ be such that $G(F)$ has no circuit of length
$\geq 2$. If $G(F)$ has no ambiguous loop, then $F$ has no cycle of length $\geq 3$.
\end{corollary}

\noindent
{\bf{Proof $-$}} Under the conditions of the statement, it is clear
that $\tau(G(F))\leq 1$. So following Theorem 2, all the cycles of $F$
are of length $\leq 2$.\cqfd

\begin{remark}
In {\emph{\cite[page 292]{R95}}}, Robert proposes to study the following
assertion: If each vertex of $G(F)$ has a loop, and if $G(F)$ has no circuit of length $\geq 2$,
then $F$ has no cycle of length $\geq 3$. This assertion is false as
showed by the following example. Let $F:\B^2\to\B^2$ be defined by:
\[
F(0,0)=(1,0),\quad F(1,0)=(0,1),\quad F(0,1)=(1,1),\quad F(1,1)=(0,0).
\]
$F$ has clearly a cycle of length $4$, but each vertex of $G(F)$ has a
loop, and $G(F)$ has no circuit of length $\geq 2$. The interaction
graph $G(F)$ is indeed the following:
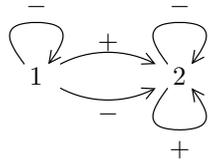
\[
\input{counterEx.pstex_t}
\]
According to the previous corrolary, the following assertion, near
that the one that Robert proposes to study, is true: If each vertex of
$G(F)$ has a loop, and if $G(F)$ has no circuit of length $\geq 2$ and no
ambiguous loop, then $F$ has no cycle of length $\geq 3$.
\end{remark}



\end{document}

%% file: ex1.pstex_t
\begin{picture}(0,0)%
\includegraphics{ex1.pstex}%
\end{picture}%
\setlength{\unitlength}{3947sp}%
\begingroup\makeatletter\ifx\SetFigFont\undefined%
\gdef\SetFigFont#1#2#3#4#5{%
  \reset@font\fontsize{#1}{#2pt}%
  \fontfamily{#3}\fontseries{#4}\fontshape{#5}%
  \selectfont}%
\fi\endgroup%
\begin{picture}(5078,1048)(3873,-5553)
\put(4050,-5063){\makebox(0,0)[b]{\smash{{\SetFigFont{10}{12.0}{\rmdefault}{\mddefault}{\updefault}{\color[rgb]{0,0,0}$i_0$}%
}}}}
\put(5850,-5063){\makebox(0,0)[b]{\smash{{\SetFigFont{10}{12.0}{\rmdefault}{\mddefault}{\updefault}{\color[rgb]{0,0,0}$i_2$}%
}}}}
\put(4950,-5063){\makebox(0,0)[b]{\smash{{\SetFigFont{10}{12.0}{\rmdefault}{\mddefault}{\updefault}{\color[rgb]{0,0,0}$i_1$}%
}}}}
\put(6750,-5063){\makebox(0,0)[b]{\smash{{\SetFigFont{10}{12.0}{\rmdefault}{\mddefault}{\updefault}{\color[rgb]{0,0,0}$i_3$}%
}}}}
\put(8775,-5063){\makebox(0,0)[b]{\smash{{\SetFigFont{10}{12.0}{\rmdefault}{\mddefault}{\updefault}{\color[rgb]{0,0,0}$i_5$}%
}}}}
\put(5401,-4988){\makebox(0,0)[b]{\smash{{\SetFigFont{10}{12.0}{\rmdefault}{\mddefault}{\updefault}{\color[rgb]{0,0,0}$+$}%
}}}}
\put(6301,-4988){\makebox(0,0)[b]{\smash{{\SetFigFont{10}{12.0}{\rmdefault}{\mddefault}{\updefault}{\color[rgb]{0,0,0}$+$}%
}}}}
\put(8251,-4988){\makebox(0,0)[b]{\smash{{\SetFigFont{10}{12.0}{\rmdefault}{\mddefault}{\updefault}{\color[rgb]{0,0,0}$+$}%
}}}}
\put(7201,-4988){\makebox(0,0)[b]{\smash{{\SetFigFont{10}{12.0}{\rmdefault}{\mddefault}{\updefault}{\color[rgb]{0,0,0}$-$}%
}}}}
\put(4501,-4988){\makebox(0,0)[b]{\smash{{\SetFigFont{10}{12.0}{\rmdefault}{\mddefault}{\updefault}{\color[rgb]{0,0,0}$-$}%
}}}}
\put(5851,-5513){\makebox(0,0)[b]{\smash{{\SetFigFont{10}{12.0}{\rmdefault}{\mddefault}{\updefault}{\color[rgb]{0,0,0}$+$}%
}}}}
\put(7726,-4613){\makebox(0,0)[b]{\smash{{\SetFigFont{10}{12.0}{\rmdefault}{\mddefault}{\updefault}{\color[rgb]{0,0,0}$-$}%
}}}}
\put(4951,-4613){\makebox(0,0)[b]{\smash{{\SetFigFont{10}{12.0}{\rmdefault}{\mddefault}{\updefault}{\color[rgb]{0,0,0}$-$}%
}}}}
\put(5851,-4613){\makebox(0,0)[b]{\smash{{\SetFigFont{10}{12.0}{\rmdefault}{\mddefault}{\updefault}{\color[rgb]{0,0,0}$-$}%
}}}}
\put(7725,-5063){\makebox(0,0)[b]{\smash{{\SetFigFont{10}{12.0}{\rmdefault}{\mddefault}{\updefault}{\color[rgb]{0,0,0}$i_4$}%
}}}}
\put(7726,-5513){\makebox(0,0)[b]{\smash{{\SetFigFont{10}{12.0}{\rmdefault}{\mddefault}{\updefault}{\color[rgb]{0,0,0}$+$}%
}}}}
\end{picture}%

%% file: ex2.pstex_t
\begin{picture}(0,0)%
\includegraphics{ex2.pstex}%
\end{picture}%
\setlength{\unitlength}{3947sp}%
\begingroup\makeatletter\ifx\SetFigFont\undefined%
\gdef\SetFigFont#1#2#3#4#5{%
  \reset@font\fontsize{#1}{#2pt}%
  \fontfamily{#3}\fontseries{#4}\fontshape{#5}%
  \selectfont}%
\fi\endgroup%
\begin{picture}(5085,1048)(3871,-5553)
\put(4050,-5063){\makebox(0,0)[b]{\smash{{\SetFigFont{10}{12.0}{\rmdefault}{\mddefault}{\updefault}{\color[rgb]{0,0,0}$i_0$}%
}}}}
\put(5850,-5063){\makebox(0,0)[b]{\smash{{\SetFigFont{10}{12.0}{\rmdefault}{\mddefault}{\updefault}{\color[rgb]{0,0,0}$i_2$}%
}}}}
\put(4950,-5063){\makebox(0,0)[b]{\smash{{\SetFigFont{10}{12.0}{\rmdefault}{\mddefault}{\updefault}{\color[rgb]{0,0,0}$i_1$}%
}}}}
\put(6750,-5063){\makebox(0,0)[b]{\smash{{\SetFigFont{10}{12.0}{\rmdefault}{\mddefault}{\updefault}{\color[rgb]{0,0,0}$i_3$}%
}}}}
\put(8775,-5063){\makebox(0,0)[b]{\smash{{\SetFigFont{10}{12.0}{\rmdefault}{\mddefault}{\updefault}{\color[rgb]{0,0,0}$i_5$}%
}}}}
\put(5401,-4988){\makebox(0,0)[b]{\smash{{\SetFigFont{10}{12.0}{\rmdefault}{\mddefault}{\updefault}{\color[rgb]{0,0,0}$+$}%
}}}}
\put(6301,-4988){\makebox(0,0)[b]{\smash{{\SetFigFont{10}{12.0}{\rmdefault}{\mddefault}{\updefault}{\color[rgb]{0,0,0}$+$}%
}}}}
\put(8251,-4988){\makebox(0,0)[b]{\smash{{\SetFigFont{10}{12.0}{\rmdefault}{\mddefault}{\updefault}{\color[rgb]{0,0,0}$+$}%
}}}}
\put(7201,-4988){\makebox(0,0)[b]{\smash{{\SetFigFont{10}{12.0}{\rmdefault}{\mddefault}{\updefault}{\color[rgb]{0,0,0}$-$}%
}}}}
\put(4501,-4988){\makebox(0,0)[b]{\smash{{\SetFigFont{10}{12.0}{\rmdefault}{\mddefault}{\updefault}{\color[rgb]{0,0,0}$-$}%
}}}}
\put(5851,-5513){\makebox(0,0)[b]{\smash{{\SetFigFont{10}{12.0}{\rmdefault}{\mddefault}{\updefault}{\color[rgb]{0,0,0}$+$}%
}}}}
\put(5851,-4613){\makebox(0,0)[b]{\smash{{\SetFigFont{10}{12.0}{\rmdefault}{\mddefault}{\updefault}{\color[rgb]{0,0,0}$-$}%
}}}}
\put(7726,-4613){\makebox(0,0)[b]{\smash{{\SetFigFont{10}{12.0}{\rmdefault}{\mddefault}{\updefault}{\color[rgb]{0,0,0}$-$}%
}}}}
\put(4951,-4613){\makebox(0,0)[b]{\smash{{\SetFigFont{10}{12.0}{\rmdefault}{\mddefault}{\updefault}{\color[rgb]{0,0,0}$-$}%
}}}}
\put(4051,-4613){\makebox(0,0)[b]{\smash{{\SetFigFont{10}{12.0}{\rmdefault}{\mddefault}{\updefault}{\color[rgb]{0,0,0}$-$}%
}}}}
\put(6751,-5513){\makebox(0,0)[b]{\smash{{\SetFigFont{10}{12.0}{\rmdefault}{\mddefault}{\updefault}{\color[rgb]{0,0,0}$+$}%
}}}}
\put(8776,-4613){\makebox(0,0)[b]{\smash{{\SetFigFont{10}{12.0}{\rmdefault}{\mddefault}{\updefault}{\color[rgb]{0,0,0}$-$}%
}}}}
\put(7725,-5063){\makebox(0,0)[b]{\smash{{\SetFigFont{10}{12.0}{\rmdefault}{\mddefault}{\updefault}{\color[rgb]{0,0,0}$i_4$}%
}}}}
\put(7726,-5513){\makebox(0,0)[b]{\smash{{\SetFigFont{10}{12.0}{\rmdefault}{\mddefault}{\updefault}{\color[rgb]{0,0,0}$+$}%
}}}}
\end{picture}%

%% file: ex3.pstex_t
\begin{picture}(0,0)%
\includegraphics{ex3.pstex}%
\end{picture}%
\setlength{\unitlength}{3947sp}%
\begingroup\makeatletter\ifx\SetFigFont\undefined%
\gdef\SetFigFont#1#2#3#4#5{%
  \reset@font\fontsize{#1}{#2pt}%
  \fontfamily{#3}\fontseries{#4}\fontshape{#5}%
  \selectfont}%
\fi\endgroup%
\begin{picture}(5085,1048)(3871,-5553)
\put(4050,-5063){\makebox(0,0)[b]{\smash{{\SetFigFont{10}{12.0}{\rmdefault}{\mddefault}{\updefault}{\color[rgb]{0,0,0}$i_0$}%
}}}}
\put(5850,-5063){\makebox(0,0)[b]{\smash{{\SetFigFont{10}{12.0}{\rmdefault}{\mddefault}{\updefault}{\color[rgb]{0,0,0}$i_2$}%
}}}}
\put(4950,-5063){\makebox(0,0)[b]{\smash{{\SetFigFont{10}{12.0}{\rmdefault}{\mddefault}{\updefault}{\color[rgb]{0,0,0}$i_1$}%
}}}}
\put(6750,-5063){\makebox(0,0)[b]{\smash{{\SetFigFont{10}{12.0}{\rmdefault}{\mddefault}{\updefault}{\color[rgb]{0,0,0}$i_3$}%
}}}}
\put(8775,-5063){\makebox(0,0)[b]{\smash{{\SetFigFont{10}{12.0}{\rmdefault}{\mddefault}{\updefault}{\color[rgb]{0,0,0}$i_5$}%
}}}}
\put(5401,-4988){\makebox(0,0)[b]{\smash{{\SetFigFont{10}{12.0}{\rmdefault}{\mddefault}{\updefault}{\color[rgb]{0,0,0}$+$}%
}}}}
\put(6301,-4988){\makebox(0,0)[b]{\smash{{\SetFigFont{10}{12.0}{\rmdefault}{\mddefault}{\updefault}{\color[rgb]{0,0,0}$+$}%
}}}}
\put(8251,-4988){\makebox(0,0)[b]{\smash{{\SetFigFont{10}{12.0}{\rmdefault}{\mddefault}{\updefault}{\color[rgb]{0,0,0}$+$}%
}}}}
\put(7201,-4988){\makebox(0,0)[b]{\smash{{\SetFigFont{10}{12.0}{\rmdefault}{\mddefault}{\updefault}{\color[rgb]{0,0,0}$-$}%
}}}}
\put(4501,-4988){\makebox(0,0)[b]{\smash{{\SetFigFont{10}{12.0}{\rmdefault}{\mddefault}{\updefault}{\color[rgb]{0,0,0}$-$}%
}}}}
\put(5851,-4613){\makebox(0,0)[b]{\smash{{\SetFigFont{10}{12.0}{\rmdefault}{\mddefault}{\updefault}{\color[rgb]{0,0,0}$-$}%
}}}}
\put(7726,-4613){\makebox(0,0)[b]{\smash{{\SetFigFont{10}{12.0}{\rmdefault}{\mddefault}{\updefault}{\color[rgb]{0,0,0}$-$}%
}}}}
\put(4951,-4613){\makebox(0,0)[b]{\smash{{\SetFigFont{10}{12.0}{\rmdefault}{\mddefault}{\updefault}{\color[rgb]{0,0,0}$-$}%
}}}}
\put(6751,-5513){\makebox(0,0)[b]{\smash{{\SetFigFont{10}{12.0}{\rmdefault}{\mddefault}{\updefault}{\color[rgb]{0,0,0}$+$}%
}}}}
\put(8776,-4613){\makebox(0,0)[b]{\smash{{\SetFigFont{10}{12.0}{\rmdefault}{\mddefault}{\updefault}{\color[rgb]{0,0,0}$-$}%
}}}}
\put(4051,-5513){\makebox(0,0)[b]{\smash{{\SetFigFont{10}{12.0}{\rmdefault}{\mddefault}{\updefault}{\color[rgb]{0,0,0}$+$}%
}}}}
\put(4051,-4613){\makebox(0,0)[b]{\smash{{\SetFigFont{10}{12.0}{\rmdefault}{\mddefault}{\updefault}{\color[rgb]{0,0,0}$-$}%
}}}}
\put(5851,-5513){\makebox(0,0)[b]{\smash{{\SetFigFont{10}{12.0}{\rmdefault}{\mddefault}{\updefault}{\color[rgb]{0,0,0}$+$}%
}}}}
\put(7725,-5063){\makebox(0,0)[b]{\smash{{\SetFigFont{10}{12.0}{\rmdefault}{\mddefault}{\updefault}{\color[rgb]{0,0,0}$i_4$}%
}}}}
\put(7726,-5513){\makebox(0,0)[b]{\smash{{\SetFigFont{10}{12.0}{\rmdefault}{\mddefault}{\updefault}{\color[rgb]{0,0,0}$+$}%
}}}}
\end{picture}%

%% file: counterEx.pstex_t
\begin{picture}(0,0)%
\includegraphics{counterEx.pstex}%
\end{picture}%
\setlength{\unitlength}{3947sp}%
\begingroup\makeatletter\ifx\SetFigFont\undefined%
\gdef\SetFigFont#1#2#3#4#5{%
  \reset@font\fontsize{#1}{#2pt}%
  \fontfamily{#3}\fontseries{#4}\fontshape{#5}%
  \selectfont}%
\fi\endgroup%
\begin{picture}(1260,1048)(3871,-5553)
\put(4050,-5063){\makebox(0,0)[b]{\smash{{\SetFigFont{10}{12.0}{\rmdefault}{\mddefault}{\updefault}{\color[rgb]{0,0,0}$1$}%
}}}}
\put(4950,-5063){\makebox(0,0)[b]{\smash{{\SetFigFont{10}{12.0}{\rmdefault}{\mddefault}{\updefault}{\color[rgb]{0,0,0}$2$}%
}}}}
\put(4951,-5513){\makebox(0,0)[b]{\smash{{\SetFigFont{10}{12.0}{\rmdefault}{\mddefault}{\updefault}{\color[rgb]{0,0,0}$+$}%
}}}}
\put(4051,-4613){\makebox(0,0)[b]{\smash{{\SetFigFont{10}{12.0}{\rmdefault}{\mddefault}{\updefault}{\color[rgb]{0,0,0}$-$}%
}}}}
\put(4501,-4838){\makebox(0,0)[b]{\smash{{\SetFigFont{10}{12.0}{\rmdefault}{\mddefault}{\updefault}{\color[rgb]{0,0,0}$+$}%
}}}}
\put(4951,-4613){\makebox(0,0)[b]{\smash{{\SetFigFont{10}{12.0}{\rmdefault}{\mddefault}{\updefault}{\color[rgb]{0,0,0}$-$}%
}}}}
\put(4501,-5288){\makebox(0,0)[b]{\smash{{\SetFigFont{10}{12.0}{\rmdefault}{\mddefault}{\updefault}{\color[rgb]{0,0,0}$-$}%
}}}}
\end{picture}%